\documentstyle[aps,prb,preprint]{revtex}
\tightenlines
\title{Structure and properties of a novel fulleride $Sm_6C_{60}$}
\author{X. H. Chen, Z. S. Liu and S. Y. Li}
\address{ Structural Research Laboratory and Department of Physics, 
University of Science and Technology of China,  Hefei, Anhui 230026, 
P. R. China}
\author{H. C. Dam, and Y. Iwasa}
\address{Japan Advanced Institute of Science and Technology\\
Tatsunokuchi, Ishikawa 923-1292,  Japan}
\date{March 12, 1999}
\begin{document}
\maketitle

\begin{abstract}
\noindent
A novel fulleride $Sm_6C_{60}$ has been synthesized using high
temperature solid state reaction. The Rietveld refinement on high
resolution synchrotron X-ray powder diffraction data shows that 
$Sm_6C_{60}$ is isostructural with body-centered cubic $A_6C_{60}$ 
(A=K, Ba). Raman spectrum of $Sm_6C_{60}$ is similar to that of 
$Ba_6C_{60}$, and the frequencies of two $A_g$ modes in $Sm_6C_{60}$ 
are nearly the same as that of $Ba_6C_{60}$, suggesting that Sm 
is divalent and hybridization between $C_{60}$ molecules and the 
Sm atom could exist in $Sm_6C_{60}$. Resistivity measurement 
shows a weak T-linear behavior above 180 K, the transport at low
temperature is mainly dominated by granular-metal theory.

\end{abstract}
 
\vskip 25 pt


\vskip 25 pt

Alkali intercalation into the $C_{60}$ host lattice is a successful
technique for synthesizing new fullerides.\cite{haddon} Hereafter, 
an extensive research has been concentrated on the intercalation of 
a wide variety of atoms or molecules in the $C_{60}$ solids. 
Intercalation of alkali metals in the $C_{60}$ solids yields various
 structural compounds $A_xC_{60}$ (x=1, 3, 4, 6) with different physical
 property.\cite{stephens,hebard,stephens1,fleming,zhou} Among 
these, the superconducting compounds, $A_3C_{60}$, has attracted
considerable interest.\cite{hebard} In this system, the fcc lattice parameter 
and band filling are tunable by changing the intercalants, $T_c$ 
goes up with increasing lattice parameter,\cite{fleming1} while it 
rapidly decreases when the nominal valence ( n ) shifts from the 
half-filling n=3 of $t_{1u}$ band.\cite{yildirim} 

The study was then extended to the alkali-earth series of $AE_xC_{60}$, 
\cite{kortan,kortan1,kortan2,kortan3} the electrons are introduced by 
intercalation of alkali-earth metals into the next lowest unoccupied 
molecular orbital with $t_{1g}$ symmetry. The superconducting 
compounds with various structures and critical temperatures were 
prepared ($Ca_5C_{60}$, $Ba_4C_{60}$, $Sr_4C_{60}$).\cite{kortan,baenitz} 
Such tolerance for the $C_{60}$ molecular valence in $t_{1g}$ 
superconductors makes a striking contrast with the strict constraint 
for the valence state in the case of $t_{1u}$ superconductors.

The rare-earth metals were also intercalated into the $C_{60}$
solids, but only one phase $RE_{2.75}C_{60}$ (RE=Yb, Sm) was 
discovered so far.\cite{ozdas,chen} In this system, the basic 
structure of $RE_{2.75}C_{60}$ is face-centered cubic, but 
the cation vacancy ordering in tetrahedral sites leads to a 
superstructure accompanied with slight lattice deformation from 
cubic to orthorhombic. The RE cations occupying the octahedral sites 
experience off-center displacements since one out of every eight 
tetrahedral sites in the subcell is vacant. Such vacancy-ordered 
structure of $RE_{2.75}C_{60}$ can be understood within a simple 
electrostatic energy model.\cite{rabe} In this paper, we report 
synthesis and structure of a novel Sm-intercalated compound 
$Sm_6C_{60}$, which is isostructural with bcc $A_6C_{60}$ 
(A=K, Ba).\cite{zhou,kortan1} It suggests that the 
cation vacancy ordering could not exist in the highly doped 
$C_{60}$ with rare earth metals. Raman scattering indicates that 
the positions of the two $A_g$ modes in $Sm_6C_{60}$ are the same 
as that of $Ba_6C_{60}$, suggesting that the Sm is divalent and 
hybridization between $C_{60}$ and Sm atoms could exist in 
$Sm_6C_{60}$. The highly Sm-doped $Sm_6C_{60}$ was confirmed 
to be metallic by resistivity measurement.

Samples of $Sm_6C_{60}$ was synthesized by  reacting stoichiometric  
amount of powers of Sm and $C_{60}$. A quartz tube with mixed powder 
inside was sealed under high vacuum of about  $2\times 10^{-6}$ torr. 
The sample of $Sm_6C_{60}$ were calcined at 550 $^oC$ for 216 hours 
with intermediate grindings of two times. X-ray diffraction(XRD) 
measurements were carried out with synchrotron radiation at the 
photon Factory of the National Laboratory for High Energy Physics 
(KEK-PF, Tsukuba). The synchrotron beam was monochromatized to 
0.8500 \AA. Rietveld refinements for the XRD patterns were carried 
out using Fullprof-98. X-ray diffraction showed that all samples were 
single phase, which is also confirmed by the single peak feature of 
the pentagonal pinch $A_g(2)$ mode in the Raman spectra.

Raman scattering experiments were carried out using the 632.8 nm line 
of a He-Ne laser in the Brewster angle backscattering geometry. The 
scattering light was detected with a Dilor xy multichannel spectrometer 
using a spectral resolution of 3 $cm^{-1}$. In order to obtain good 
Raman spectra, the samples were ground and pressed into pellets with 
pressure of about 20 $kg/cm^2$, which were sealed in Pyrex tubes under 
a high vacuum of $10^{-6}$ torr.

Resistivity measurements were carried out using four-probe method. 
Electrical contacts of less than 2 $\Omega$ resistance were established
by silver paste. Preparation of samples and electrical contacts was 
carried out in a controlled argon glove box where the oxygen and water 
vapor levels were maintained below a few parts per million.

Figure 1 shows the XRD pattern of the $Sm_6C_{60}$ sample. The pattern
is similar to that for the well-known bcc structure found in $A_6C_{60}$
 (A=K, Rb, Cs, and Ba).\cite{zhou,kortan1} All the observed peaks are 
indexed with bcc lattice of a=10.890 \AA. The diffraction pattern of 
$Sm_6C_{60}$ samples fits well to a single body-centered cubic structure.
 We have carried out a Rietveld refinement of the structure using the 
Fullprof-98. The solid line in Fig.1 shows fitted data to a model of the 
bcc structure (space group $I_m$$_3^-$).  Refinement of 220 peaks yielded 
an intensity R factor of $R_{wp}$=9.9\%
and $R_p$=9.4\%, and the following coordinates: C1 at 0.0660, 0.0, 0.3201;
C2 at 0.1320, 0.1070, 0.2796; C3 at 0.0660, 0.2140, 0.2392; and Sm at 0.0, 
0.5, 0.2749. Mean-square thermal amplitudes of the isotropic Debye-Waller 
factors were refined to 0.016 and 0.103 $\AA ^2$ for C and Sm, respectively, 
and all samarium sites were found to be occupied. The
refined lattice constant, a=10.890 \AA, however, is significantly 
smaller than measured values for $K_6C_{60}$ and $Cs_6C_{60}$,\cite{zhou} $Ba_6C_{60}$,\cite{kortan1} and $Rb_6C_{60}$,\cite{zhu}  and 
suggests a 9.43 $\AA$ nearest-neighbor separation for the molecules. 
The atomic coordinates of C1, C2, C3 and Sm are nearly the same as 
that of $A_6C_{60}$ (A=K, Rb, Cs, and Ba).\cite{zhou,kortan1,zhu} 

For the low Sm-doped $C_{60}$,  $Sm_{2.75}C_{60}$ is isostructural 
with $Yb_{2.75}C_{60}$,\cite{chen} and different from the fcc $A_3C_{60}$ 
(A=K, Rb).\cite{stephens1} In this structure, the fcc-based subcell 
of four $C_{60}$ molecules has Sm cations occupying all four 
octehedral (O) interstitial sites but only seven of the eight 
tetrahedral (T) sites. The unoccupied site alternates between 
adjacent T sites in each direction, leading to an unit-cell 
doubling with eight ordered vacancies. The O-site Sm cations 
are displaced from the centers of their interstices towards the 
nearest neighbor vacancy, effectively reducing the nearest 
neighbor coordination of $C_{60}$ anions around the O-site 
cation from six to three. Thus, in the unit cell there are three 
types of $C_{60}$ molecules. In contrast to the $RE_{2.75}C_{60}$,
 the highly Sm-doped $Sm_6C_{60}$ is isostructural 
with $A_6C_{60}$ (A=K, Rb, Cs, and Ba).\cite{zhou,kortan1,zhu} It
 suggests that further intercalation of Sm into $C_{60}$ solids 
leads to disappearance of cation-vacancy ordering and formation 
of a cation-disordered phase. The number of interstitial sites
increases from three per $C_{60}$ in a faced-centred cubic (fcc)
structure to six in the bcc structure. Additionally, the distinction 
between octahedral and tetrahedral sites is removed in the bcc 
structure, all interstitial sites becoming equivalent with distorted
tetrahedral symmetry. For the  $RE_{2.75}C_{60}$, the vacancies 
create three inequivalent types of $C_{60}$ anions. Each anion 
rotates about an internal axis to maximize the number of pentagon 
oriented towards the surronding cations. In the highly Sm-doped 
$Sm_6C_{60}$, all interstitial sites are filled by six samarium 
atoms. All $C_{60}$ balls are equivalent and 
orientationally uniform, and all the tetrahedral holes are surrounded
by two pentagons and two hexagons.

Figure 2 shows room temperature Raman spectrum for the 
polycrystalline sample of $Sm_6C_{60}$. In the spectrum, only
one peak of pentagonal pinch $A_g(2)$ mode is observed, providing
an evidence that the sample is a single phase. This agrees fairly 
well with the x-ray diffraction patterns. It is worthy to note that
the Raman spectrum of $Sm_6C_{60}$ is amazingly similar to that of 
$Ba_6C_{60}$,\cite{chen1} suggesting that the electronic states 
of $Sm_6C_{60}$ is similar to that of $Ba_6C_{60}$. The positions 
($\omega$) and halfwidths ($\gamma$) of the Raman modes observed 
are listed in Table I. For comparison, the lines for pure $C_{60}$ and 
$Ba_6C_{60}$ are included in Table I. The frequencies of the two 
$A_g$ derived modes are 505.5 and 1371 $cm^{-1}$, respectively. 
Which are different from 498.3 and 1432.8 $cm^{-1}$ observed 
for the corresponding $A_g$ modes in $Sm_{2.75}C_{60}$.\cite{chen2}
It also provides a direct evidence for an existence of new phase 
in Sm-doped $C_{60}$ system. It is seen in table I that the 
frequencies of the two $A_g$ modes are nearly the same as those 
in $Ba_6C_{60}$. It suggests that $C_{60}$ in $Sm_6C_{60}$ is 
hexavalent, being in a fair agreement with a simple expectation 
assuming that Sm cation is divalent. This is consistent with the 
results of near-edge and extended X-ray absorption fine structure in 
$Yb_{2.75}C_{60}$,\cite{citrin} and Raman scattering results of 
$Sm_{2.75}C_{60}$,\cite{chen2} in which Sm cation is comfirmed to 
be divalent.  
  
Two theoretical calculations based the local density approximation
have shown a strong hybridization between the alkaline-earth-atom 
states and the $C_{60}$ $\pi$ states.\cite{saito,erwin} Recent 
photoemission studies have also indicated the presence of a 
hybrid band with incomplete charge transfer from alkaline-earth 
metals to $C_{60}$, as a consequence of the competition between 
covalent $Ba-C_{60}$ bonding and ionic contribution.\cite{niedrig}
 The same frequency of the $A_g(2)$ mode as that in $Ba_6C_{60}$,
which is known as a sensitive probe for the degree of charge 
transfer on $C_{60}$ molecule, indicates 
that the hybridization between Sm atom and $C_{60}$ molecules could 
exist in $Sm_6C_{60}$. In fact, the size of the divalent samarium, 
1.18 \AA, placed at (0.22, 0.5, 0) requires a lattice constant of 
about 11.5 $\AA$ assuming spherically symmetric molecules. The 
relatively small lattice constant measured also may indicate strong 
orbital overlap between the neighboring molecules. This hybridization 
may play an essential role for a larger Raman downshift of the 
pentagonal pinch $A_g(2)$ mode than that expected by  the simple 
relation between Raman shift and charge transfer widely observed 
in $K_xC_{60}$.\cite{duclos,kuzmany}  

In the table I, the positions and halfwidths were obtained by fitting
the experimental data with Lorentzian line shape. It is easily seen
that the lowest frequency $H_g$ modes are splitted into several 
components. The similar behavior has been observed in single crystal
 $K_3C_{60}$ at low temperature\cite{winter} and in $Ba_xC_{60}$ 
(x=4 and 6).\cite{chen1} $H_g(2)$ mode is apparently split into five 
components in $Sm_6C_{60}$. The inset of Fig.2 shows the results of a 
line-shape analysis for $H_g(2)$ mode. It suggests that the five-fold 
degeneracy is 
completely lift. This splitting of $H_g(2)$ mode in $Sm_6C_{60}$
is unexpected since the group theoretical consideration predicts a 
splitting into two in the space group $I_m$$_3^-$ ($T_5^h$). The 
splitting might suggest a symmetry lowering which is not detected 
in the x-ray diffraction. This type of disagreement between 
microscopic spectroscopy and structural analysis was observed 
in $Rb_3C_{60}$\cite{walstedt} and $Ba_6C_{60}$.\cite{chen1} From 
Table I, we can find that the positions for most of modes are 
the same, and the similar splitting is observed in $Sm_6C_{60}$ 
and $Ba_6C_{60}$. This strongly indicates that $Sm_6C_{60}$ and 
$Ba_6C_{60}$ have a similar crystal structure and electronic states.
     
Figure 3 show the temperature dependence of resistivity for the sample 
$Sm_6C_{60}$. No superconducting transition is observed at the 
temperature down to 4.2 K. A minimum resistivity appears at about 180 K. The 
temperature coefficient of resistivity  is  positive over 180 K, while
is negative below 180 K. The data in Fig.3 are fitted by the
 following formula:
\begin{equation}
\rho =ae^{-bT^{-1/2}}+cT+d  
\end{equation}
where a, b, c, and d are the fitting parameters. It is found that 
data  can be well fitted over the whole temperature range of 4.2-300 K by 
the formula. The first item in the formula is written according
to the granular metal theory, in which the $ln \rho$ is proportional 
to the $T^{-1/2}$. The linear-T item is used to fit the metallic 
behavior at the high temperature. The inset of Fig.3 plots the same 
data as $ln \rho$ vs $T^{-1/2}$, and almost shows a linear relation  below
180 K. It suggests that the transport is dominated by the granular-metal
theory. Stepniak et al. have reported that the thin film samples of 
$Rb_xC_{60}$ can be described within the framework of granular metal 
theory.\cite{stepniak}  The sample $Sm_6C_{60}$ shows a weak localization 
behavior at low temperature and the ratio $\rho(4.2 K)/\rho(290 K)$ is 
only 1.7. In addition, the transport at low temperature can be explained 
by the granular metal theory. These results indicate that the sample
$Sm_6C_{60}$ is metallic. This is in apparent contradiction to 
the insulating electronic structure if the charge complete transfer 
from the rare-earth divalent Sm atom to $C_{60}$ molecules could
result in the full occupation of the $t_{1g}$ band. The metallic 
behavior of $Sm_6C_{60}$ suggests that there could exist a 
hybridization between rare-earth Sm $f$ orbitals and $C_{60}$ $p$ 
$\pi$ orbitals, being similar to the case of $Ba_6C_{60}$. In the 
case of $Ba_6C_{60}$, the theoretical calculation based on the 
local-density approximation shows that the hybridization between 
the Ba atom s states and the $C_{60}$ $\pi$ states is essential 
for the metallic electronic structure.\cite{saito}  In addition, 
the same Raman shift of the $A_g(2)$ pinch mode for $Sm_6C_{60}$ 
and $Ba_6C_{60}$ also provides a direct evidence for a hybridization 
between Sm atoms and $C_{60}$ molecules. In order to further confirm  
the hybridization in $Sm_6C_{60}$, a theoretical calculation or 
photoemission study is necessary.

In summary, we synthesized a novel fulleride $Sm_6C_{60}$, which
was characterized by X-ray diffraction, Raman scattering and
resistivity measurement. $Sm_6C_{60}$ adopts body-centered cubic
structure, being similar to that of $A_6C_{60}$ (A=K, Rb, Ba). 
Raman spectrum of $Sm_6C_{60}$ is strikingly similar to that 
of $Ba_6C_{60}$. The same Raman shift of the $A_g(2)$ pinch mode 
in $Sm_6C_{60}$ and $Ba_6C_{60}$ suggests that Sm is divalent and
there exists a hybridization between rare-earth Sm atom and the 
$C_{60}$ molecules, which could be responsible for the metallic 
behavior of $Sm_6C_{60}$ observed by resistivity measurement.
A weak T-linear metallic behavior is observed down to 180 K, the 
resistivity data at low temperature can be explained by 
the granular-metal theory.

\section*{acknowlegments}

This work is supported by Grant from Natural Science Foundation
of China.

\begin{table}
\caption{ Positions and linewidths  (in parentheses)  for the Raman modes  in $C_{60}$ and $Ba_6C_{60}$   }
\vskip 10 pt
\renewcommand{\arraystretch}{0.93}
\begin{tabular}{c c c c } 
         &  $C_{60}$  &  $Ba_6C_{60}$  &  $Sm_6C_{60}$  \\
$I_h$ mode & $\omega$ ( $\gamma$ ) & $\omega$ ( $\gamma$ ) &  
$\omega$ ( $\gamma$ )  \\
        & ( $cm^{-1}$ ) & ( $cm^{-1}$ ) & ( $cm^{-1}$ )  \\ \hline
$A_g(1)$  &  493  &  506.5  ( 5.0 )   &  505.5  ( 3.3 )  \\
$A_g(2)$  &  1469 &  1372.5 ( 12.1 )  &  1371.0 ( 7.0 )  \\
$H_g(1)$  &  270  &  274.5  ( 5.2 )   &  263.4  ( 7.8 )  \\
          &       &  281.8  ( 2.6 )   &  280.0  ( 13.9 ) \\
          &       &                   &  291.8  ( 3.7 )  \\
$H_g(2)$  &  431  &  385.6 ( 4.4 )    &  367.2  ( 7.1 )  \\
          &       &  405.8 ( 2.2 )    &  391.6  ( 5.1 )  \\
          &       &  415.6 ( 2.4 )    &  400.8  ( 4.1 )  \\
          &       &  428   ( 16.8 )   &  415.9  ( 28.2 ) \\
          &       &  438.8 ( 2.8 )    &  426.8  ( 6.4 )  \\
$H_g(3)$  & 709   &  585.2 ( 4.8 )    &  587.5  ( 28.0 ) \\
          &       &  602.1 ( 5.2 )    &  603.2  ( 10.8 ) \\
          &       &  622.3 ( 3.7 )    &  638.0  ( 32.0 ) \\
          &       &  651.8 ( 12.0 )   &                  \\
$H_g(4)$  & 773   &  732.5 ( 8.5 )    &  740.0  ( 9.0 )  \\
$H_g(5)$  & 1099  &  1082 ( 6.0 )     &  1084.0 ( 12.0 ) \\
          &       &                   &  1110.2 ( 10.8 ) \\
$H_g(6)$  & 1248  &  1224 ( 26 )      &  1210.5 ( 6.5 )  \\
          &       &                   &  1227.2 ( 10.3 ) \\
$H_g(7)$  & 1426  &                   &                  \\
$H_g(8)$  & 1573  &  1437 ( 25.0 )    &  1440.5 ( 20.2 ) \\
\end{tabular}
\end{table}

\newpage
\noindent
{\bf FIGURE CAPTIONS} \\

\noindent
Figure 1:

X-ray diffraction pattern of the sample $Sm_6C_{60}$ collected with synchrotron 
radiation. The synchrotron beam was monochromatized to 0.8500 \AA. The crosses 
are experimental points and the solid line is a Rietveld fit to the model 
$Sm_6C_{60}$ in the space group $I_m$$_3^-$. The allowed reflection positions 
are denoted by ticks.\\

\noindent
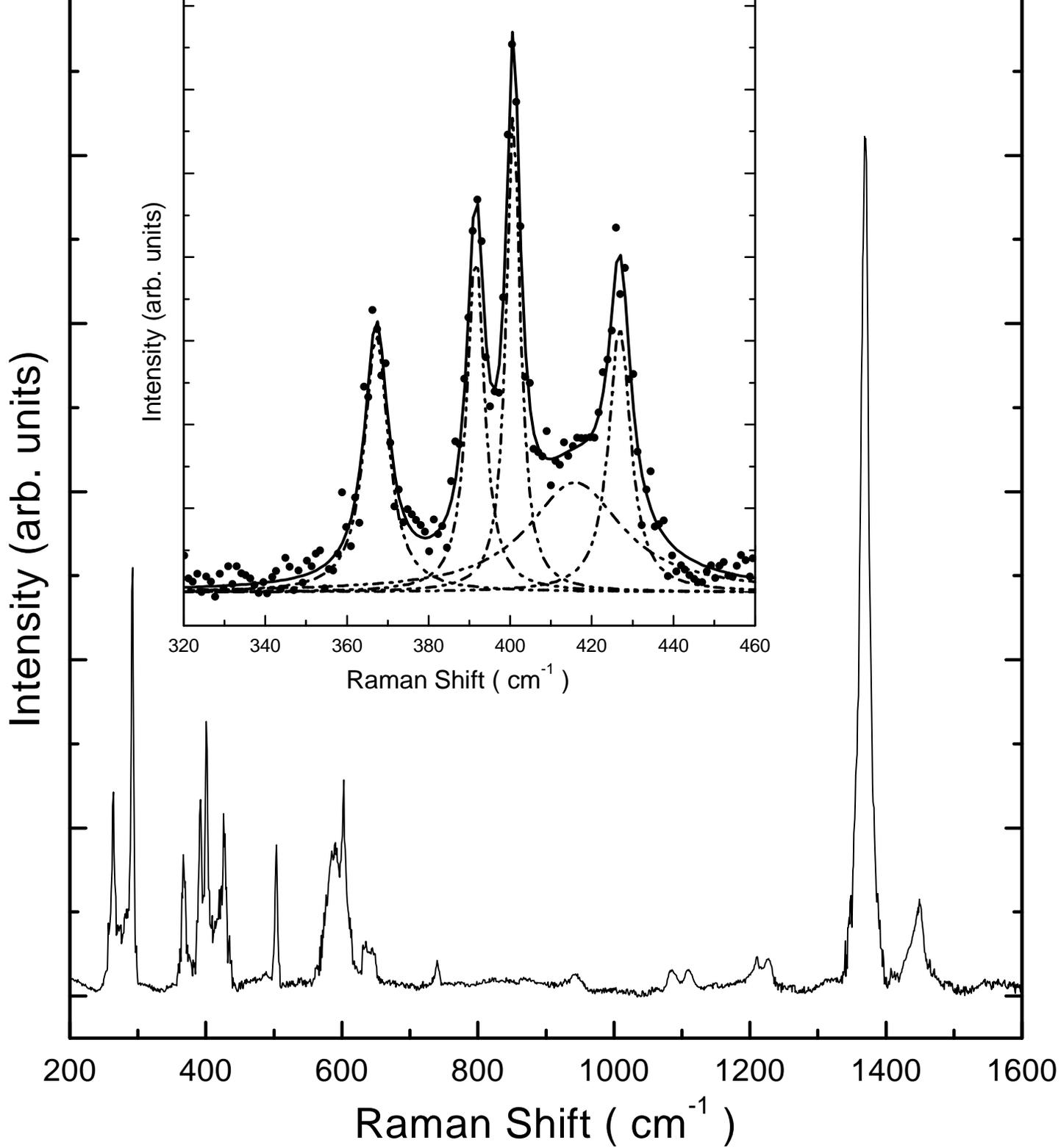
Figure 2:

Room temperature Raman spectrum of $Sm_6C_{60}$.  The results of a line-shape 
analysis for $H_g(2)$ mode are shown (inset). The dash lines are computer 
fits for the individual components, which add up to the full 
line on the top of the experimental results.\\

\noindent
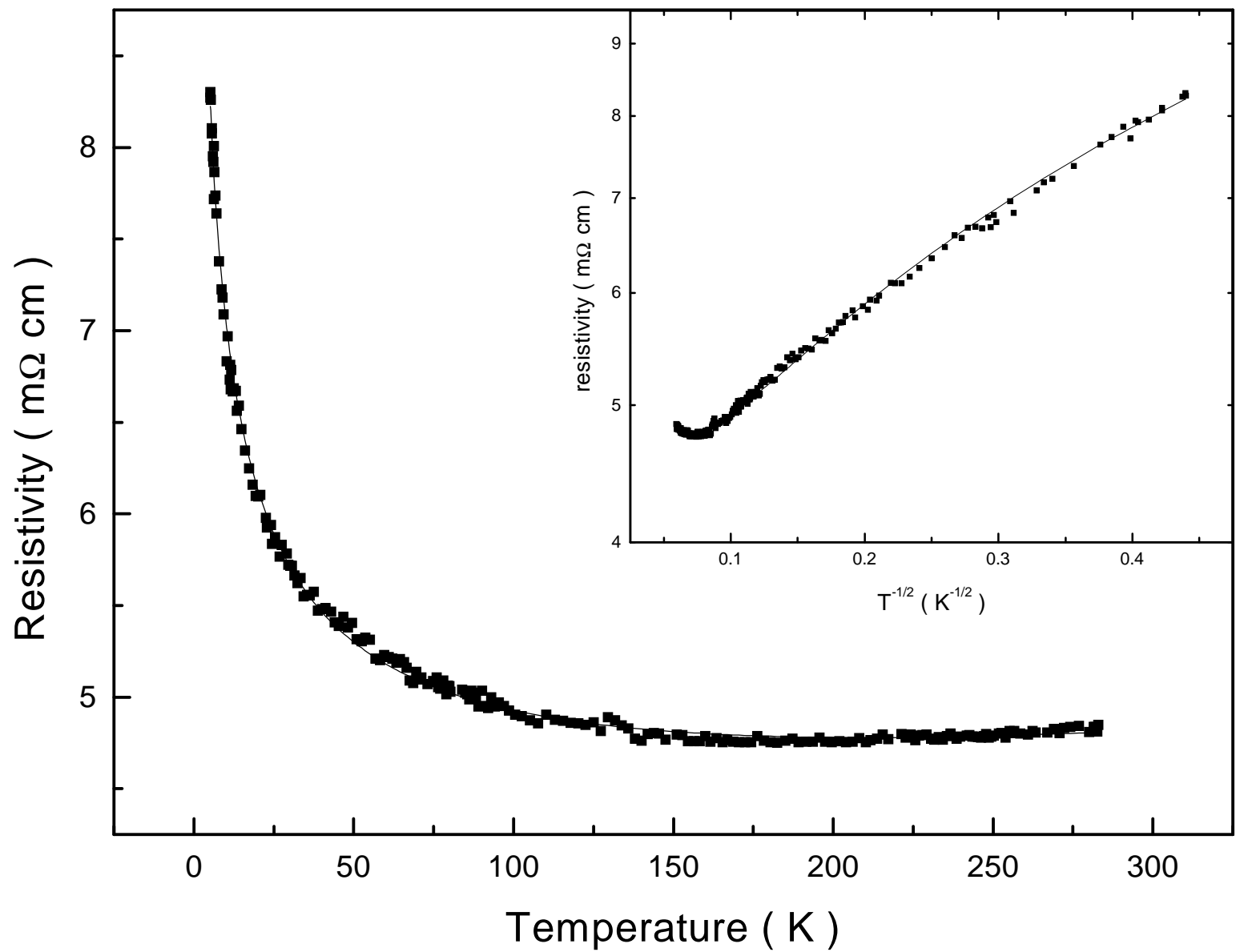
Figure 3:

The temperature dependence of resistivity for the polycrystalline 
$Sm_6C_{60}$. Inset plots the same data as $ln\rho$ vs $T^{-1/2}$.
A linear relation would be expected if the charge transport did
follow the granular-metal theory.


\begin{references}

\bibitem{haddon}
R.C. Haddon, A.F. Hebard, M.J. Rosseinsky, D.W. Murphy, S.J. Duclos,
K.B. Lyons, B. Miller, J.M. Rosamilia, R.M. Fleming, A.R. Kortan,
S.H. Glarum, A.V. Makhija, A.J. Muller, R.H. Eick, S.M. Zahurak, 
R. Tycko, G. Dabbagh, and F.A. Thiel, Nature {\bf350}, 320(1991).
\bibitem{stephens}
P.W. Stephens, G. Bortel, G. Faigel, M. Tegze, A. Janossy, S. Pekker,
G. Oszlanyki, L. Forro, Nature {\bf370}, 636(1994).
\bibitem{hebard}
A.F. Hebard, M.J. Rosseinsky, R.C. Haddon, D.W. Murphy, S.H. Glarum, 
T.T.M. Palstra, A.P. Ramirez, and A.R. Kortan, Nature {\bf350}, 
600(1991). 
\bibitem{stephens1}
P.W. Stephens, L. Mihaly, P.L. Lee, R.L. Wheten, S.M. Huang, R.B. Kaner,
F. Diederich, and K. Holczer, Nature {\bf351}, 632(1991).
\bibitem{fleming}
R.M. Fleming, M.J. Rosseinsky, A.P. Ramirez, D.W. Murphy, J.C. Tully,
R.C. Haddon, T. Siegrist, R. Tycko, H. Glarum, P. Marsh, G. Dabbagh,
S.M. Zahurak, A.V. Makhija, and C. Hampton, Nature {\bf352}, 701(1991).
\bibitem{zhou}
O. Zhou, J.E. Fisher, N. Coustel, S. Kycia, Q. Zhu, A.R. McGhie, 
W.J. Romanow, J.P.Jr. McCauley, A.B. Smith, and D.E. Cox, 
Nature {\bf351}, 462(1991).  
\bibitem{fleming1}
R.M. Fleming, A.P. Ramirez, M.J. Rosseinsky, D.W. Murphy, R.C. Haddon,
S.M. Zahurak, and A.V. Makhija, Nature {\bf352}, 787(1991).
\bibitem{yildirim}
T. Yildirim, L. Barbedette, J.E. Fisher, C.L. Lin, J. Robbert, P. Petit,
and T.T.M. Palstra, Phys. Rev. Lett. {\bf77}, 167(1996).
\bibitem{kortan}
A.R. Kortan, N. Kopylov, S. Glarum, E.M. Gyorgy, A.P. Ramirez, R.M. Fleming,
 F.A. Thiel, and R.C. Haddon, Nature {\bf355}, 529(1992).
\bibitem{kortan1}
A.R. Kortan, N. Kopylov, S. Glarum, E.M. Gyorgy, A.P. Ramirez, R.M. Fleming,
O. Zhou, F.A. Thiel, P.L. Trevor, and R.C. Haddon, Nature {\bf360}, 566(1992).   
\bibitem{kortan2}
A.R. Kortan, N. Kopylov, E. \"Ozdas, A.P. Ramirez, R.M. Fleming,
 and R.C. Haddon, Chem. Phys. Lett. {\bf233}, 501(1994).   
\bibitem{kortan3}
A.R. Kortan, N. Kopylov, R.M. Fleming, O. Zhou, F.A. Thiel, and R.C. Haddon, 
Phys. Rev. B {\bf47}, 13070(1993).
\bibitem{baenitz}
M. Baenitz, M. Heinze, K. L\"uders, H. Werner, R. Sch\"ogl, M. Weiden,
G. Sparn, and F. Steglich, Solid State Commun. 96, 539(1995).
\bibitem{ozdas}
E. \"Ozdas, A.R. Kortan, N. Kopylov, A.R. Ramirez, T. Siegrist, K.M. Rabe,
H.E. Bair, S. Schuppler, and P.H. Citrin, Nature {\bf375}, 126(1995).
\bibitem{chen}
X.H. Chen and G. Roth, Phys. Rev. B {\bf52}, 15534(1995). 
\bibitem{rabe}
K.M. Rabe and P.H. Citrin, Phys. Rev. B {\bf58}, 551(1998).
\bibitem{zhu}
Q. Zhu, O. Zhou, N. Coustel, G.B.M. Vaughan, J.P.Jr. McCauley, W.J. Romanov,
J.E. Fisher, and A.B. Smith, Science {\bf254}, 545(1991).   
\bibitem{chen1}
X.H. Chen, S. Taga, and Y. Iwasa, Phys. Rev. B  (to be published).
\bibitem{chen2}
X.H. Chen, T. Takenobu, T. Muro, H. Fudo, and Y. Iwasa, 
 submitted to Phys. Rev. B
\bibitem{citrin}
P.H. Citrin, E. \"Ozdas, S. Schuppler, A.R. Kortan, and K.B. Lyons
Phys. Rev. B {\bf56}, 5213(1997).
\bibitem{saito}
S. Saito and A. Oshiyama, Phys. Rev. Lett. {\bf71}, 121(1993).
\bibitem{erwin}
S.C. Erwin and M.R. Pederson, Phys. Rev. B {\bf47}, 8249(1993).
\bibitem{niedrig}
Th. Schedel-Niedrig, M.C. Bohm, H. Werner, J. Schulte, and R. Schlogl,
Phys. Rev. B {\bf55}, 13542(1997).
\bibitem{duclos}
S.J. Duclos, R.C. Haddon, S.H. Glarum, A.F. Hebard, and K.B. Lyons,
Science {\bf254}, 1625(1991).
\bibitem{kuzmany}
H. Kuzmany, M. Matus, B.Burger, and J. Winter, Adv. Mater. {\bf6},
731(1994).
\bibitem{winter}
J. Winter and H.Kuzmany, Phys. Rev. B {\bf53}, 655(1996).
\bibitem{walstedt}
R.E. Walstedt, D.W. Murphy, and M.J. Rosseinsky, Nature {\bf362},
611(1993).
\bibitem{stepniak}
F. Stepniak, P.J. Benning, D.M. Poirier, and J. Weaver,
Phys. Rev. B {\bf48}, 1899(1993).

\end{references}
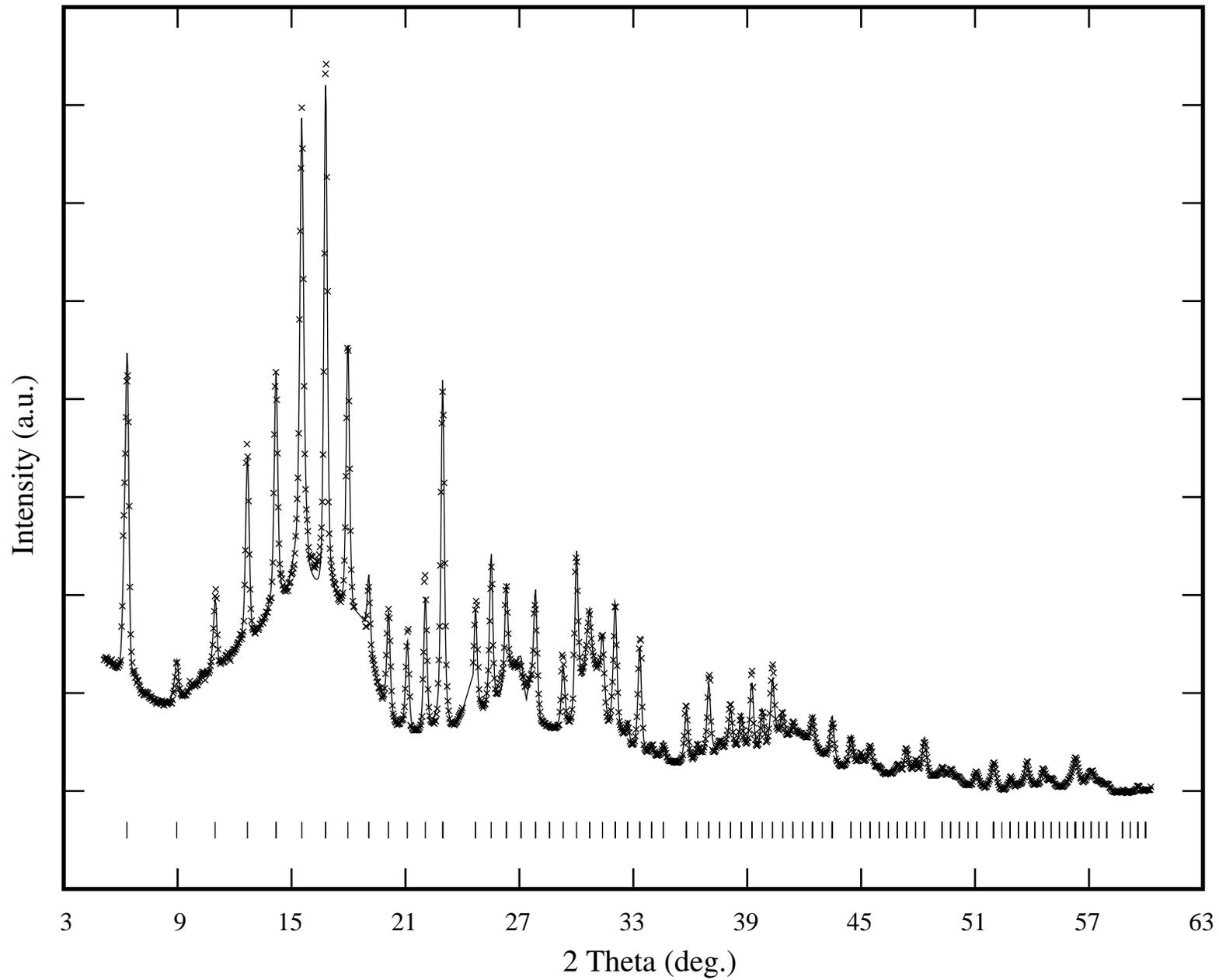
\end{document}